# THE SIZE EFFECT ON NUCLEATION PROCESS DURING SOLIDIFICATION OF METALS

A.S. Nuradinov[1], K.A. Sirenko[1], I.A. Nuradinov[1], O.V. Chystiakov[1], D.O. Derecha[1,2*]

[1]Physico-technological Institute of Metals and Alloys Natl. Acad. of Sci. of Ukraine. Acad. Vernadskoho 34/1, Kyiv, Ukraine, 03680.

[2]V.G. Baryakhtar Institute of Magnetism Natl. Acad. of Sci. of Ukraine. Acad. Vernadskoho 34, Kyiv, Ukraine, 03142.

* Corresponding author: Dmytro Derecha. E-mail: dderecha@gmail.com

**Abstract**

This work investigates the mechanisms of crystal nucleation in metal melts, in dependence on the influence of their shape and volume. Investigations were conducted on both thin flat and bulk samples prepared of low-temperature Wood's metal alloy and transparent organic media (camphene and diphenylamine). The study revealed that in flat samples, the nucleation rate is primarily influenced by the activity of the mold wall surface, which is affected by melt overheating and subcooling, while in bulk samples, supercooling and nucleation are additionally influenced by factors such as impurity activity, temperature, and density fluctuations.

**Introduction.**

The modern stage of the development of mechanical engineering is characterized by a tendency to increase the requirements for the quality of the metal products used, which is largely determined by the quality of the cast blanks at the first stage of their production. Accordingly, when obtaining cast blanks, it is assumed that metal products made of cast metal with a higher dispersion of the crystal structure, minimal liquation and uniform distribution of non-metallic inclusions have the highest level of mechanical properties. To obtain cast blanks with such a metal structure, it is necessary to ensure the maximum rate of crystal nucleation, which depends on many factors, including the volume and values of overheating and subcooling of the melt before crystallization. Therefore, many works are devoted to the problem of nucleation and growth of crystals in metals, which confirms the importance of the volume and the amount of subcooling of melts in the processes of their crystallization [1-13, 10-18].

There are theories of surface [5-7] and volume crystallization [3, 4, 8, 9], which explain the dependence of the rate of nucleation on overheating (subcooling) in different ways. In the first case, it is considered that crystallization begins at the point of contact of the melt with the surface of the crucible, solid particles or oxide film, and overheating affects the state of this surface. For example, it is assumed that solid metal is preserved in the pores of the crucible (after repeated melting of the metal in the same crucible) when overheated to some critical temperature. The solid metal is preserved due to the surface energy and the increased pressure it undergoes in the pore due to the greater dilatation coefficient of the metal than that of the oxide [6, 7]. This theory can explain the change in the crystallization mechanism under low values of superheat up to several degrees or a maximum of tens of degrees above the melting temperature. But this explanation is not suitable, for example, for aluminium, for which the hypothermia plateaus (that is, reaches a maximum and does not change) with superheats greater than 200°C [3, 4]. An alternative theory allows the micro-groupings of atoms (clusters) to exist in the melt, which are preserved under slight superheating of the melt over the liquidus and serve as nuclei during crystallization [8, 9]. But the cluster theory has also not been experimentally confirmed for pure aluminium [7]. In general, both theories found partial experimental confirmation in experiments with metals [3, 4, 7, 10-12].

Recent advances have revolutionized the understanding and control of crystallization processes. Advanced computational models, such as molecular dynamics, Monte-Carlo methods, and phase-field models, allow for precise prediction of nucleation rates and crystal morphologies, facilitating the rational design of materials with desired properties. Innovative strategies have emerged to enhance control over crystal growth kinetics and crystallographic orientations, alongside process intensification strategies like microreactors and membrane crystallization, which boost nucleation rates and crystal growth. The potential-driven growth of metal crystals from ionic liquids is also being explored. Furthermore, in-situ X-ray imaging techniques have provided significant insights into crystal formation, growth, instability, and defects during alloy solidification. Research also indicates that crystallizations often involve multiple intermediate structures, and the atomistic pathways from metastable phases to final crystals are a subject of ongoing investigation, including the concept of two-step nucleation.

**Materials and techniques.**

The direct study of the processes related to the casting of metals and their alloys is extremely complicated due to their opacity and very high flow temperatures. Therefore, in this work, the method of physical modeling was applied, utilizing low-temperature ($t_m \leq 70°C$) Wood's (12.5% Sn, 12.5% Cd, 25% Pb, 50% Bi) metal alloy without a crystallization interval and transparent organic media such as camphene ($C_{10}H_{16}$) with a crystallization interval of 3°C and diphenylamine ($C_{12}H_{11}N$) without a crystallization interval see Table 1. All model materials had high purity (~ 99%). Experiments were conducted on two types of test samples: flat samples of small thickness ($\delta = 0.2$ mm) and volume samples of larger volumes in quartz tubes (Ø 8 mm). Flat samples were used to exclude the influence of volume factors like convection, temperature, and density fluctuations on the nucleation process in organic melts due to their small layer thickness. The effect of these volumetric factors on nucleation in model media melts was studied in larger volume samples within glass tubes, where these factors are certainly present.

Table 1. – Physical properties of model materials

| Material / Properties | Wood alloy | Camphene | Diphenylamine |
|---|---|---|---|
| Density, kg/m$^3$ | 9720 | 845 | 1200 |
| Liquidus temperature, °C | 68 | 45 | 53 |
| Solidus temperature, °C | 68 | 42 | 53 |
| Crystallization interval, °C | 0 | 3 | 0 |

For this purpose, an experimental setup (flat model) was prepared, in which samples in the form of flat layers from the melts of the studied media (Fig. 1) were obtained. The flat layers of melts were obtained by using round glass plates 1 and 2, which were fixed in parallel in the bodies 3 of the model and pressed against each other with the help of a cover 4 with a threaded connection. The preselected thickness of the melt layer was provided by installing a spacer 5 of aluminium foil between the glass plates. The hermiticity of the internal cavity against the penetration of the coolant (water) was ensured by using a heat-resistant silicone sealant during its assembly.

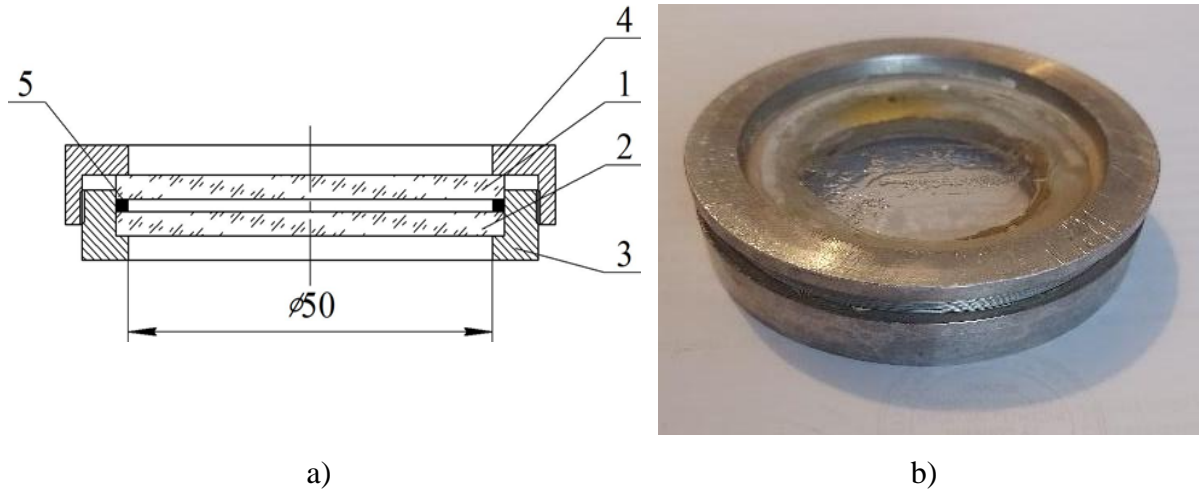

a)                                                                             b)

Fig. 1. Experimental setup. Scheme (a) and photo (b).

The experiments were carried out as follows: a flat model with an experimental medium was placed in a cuvette with transparent sides, into which a heat carrier (water) with the required temperature (measurement accuracy ± 0.1°C) was supplied from a thermostat by a circulation pump. At the beginning of the experiment, the environment in the model was melted by setting the temperature of the heat carrier necessary for this, and kept at this temperature for 10 minutes. Then, from another thermostat, a cooling liquid (water) was supplied to the cuvette with a temperature that corresponded to a certain subcooling for a specific environment and was sufficient for the formation of a nucleus. The small thickness of the melt layer and the transparency of the media made it possible to visually observe the crystallization process over time. The number of crystals that appeared at each moment of their fixation was counted according to the corresponding photographs. For different experimental conditions (that is, depending on the temperatures of superheating and subcooling of the melt), the rate of crystal nucleation was calculated based on the number of crystals that appeared on the observed surface area of the model over time ($I$) according to the following formula:

$$I = \frac{1}{V} \cdot \frac{dn}{d\tau}, \qquad (1)$$

where $dn$ is the number of crystals on the surface area of the model that were formed during the corresponding time $d\tau$; $V$ is the volume of melt in the model.

For studies of the volumetric factors' effect on the nucleation process in melts of model environments samples of larger volumes in quartz tubes (Ø 8 mm) were used, and a special experimental setup was created (Fig. 2).

The research was conducted according to the following methodology. First, three test samples of the same volume were prepared from each alloy in quartz tubes ($V \leq 18 \times 10^{-6}$ m³). To ensure the absolute identity of the melting and crystallization conditions of the model alloys, all three tubes with test samples 1 were simultaneously placed using a special fixture (cassette) 2 in a cuvette 3, into which a coolant (water) was supplied from the thermostat 4 by a circulation pump with the required temperature (sufficient to melt and superheat the model alloy to a certain level). After exposure of the samples in an overheated state (10 min), their cooling began at a rate of 2 °/min.

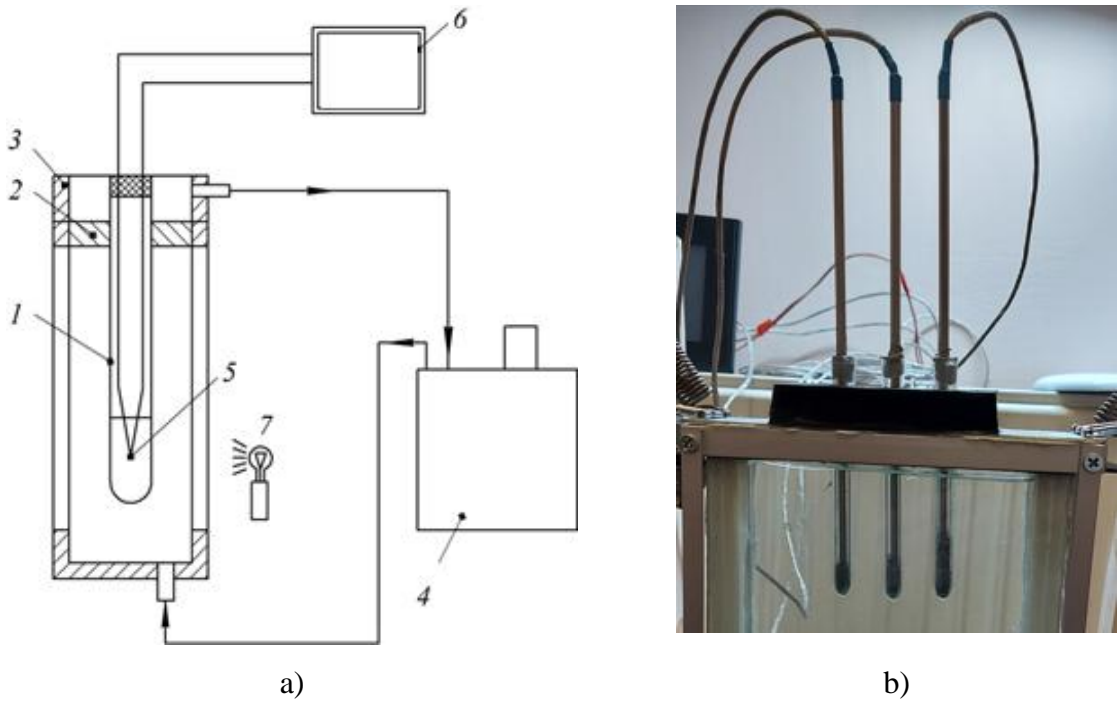

a)            b)

Fig. 2. Scheme of the experimental setup (a) and photo (b).

The temperature measurements of crystal formation in the samples the thermocouples 5 were installed in test tubes with experimental medium 1, the signal from which in the form of absolute temperature values was fixed by digital potentiometer 6 and stored on the memory card. These temperature values were converted into the melts cooling curves of the experimental environments. According to the characteristic signs on the temperature curve of the cooling of the model environment (that is, when plateau appear on it due to the release of heat of crystallization), the amount of supercooling of the melt, at which crystal nucleation occurs, was determined. The transparency of the organic model media, the coolant, and the faces of the cuvette made it possible to observe the nucleation of crystals in them also visually. To ensure the improvement of visual observation of the crystallization processes of the model media in the test tubes, the cuvette 3 was illuminated with light from the lamp 7.

**Results and discussion.**

As a result of research, it was established that the change in the rate of formation of nucleus (*I*) in time (*t*) in flat samples of transparent model media, calculated according to formula (1), for all conditions of the experiment had an extreme character (Fig. 3). In this figure, from the example of camphene, we can see that *I* initially grow rapidly and, having reached a maximum, falls to zero. At the same time, with an increase in the superheat temperature of the melt ($\Delta t^+$) maximum value $I_{max}$ decreases and shifts to the right, i.e., $I_{max}$ is achieved in a longer time. We observed a similar nature of the change in the rate of formation of nuclei over time in flat samples with an increase in their supercooling ($\Delta t^-$) for the same level of their overheating ($\Delta t^+$). All this indicates a non-stationary mode of formation of nuclei in these environments.

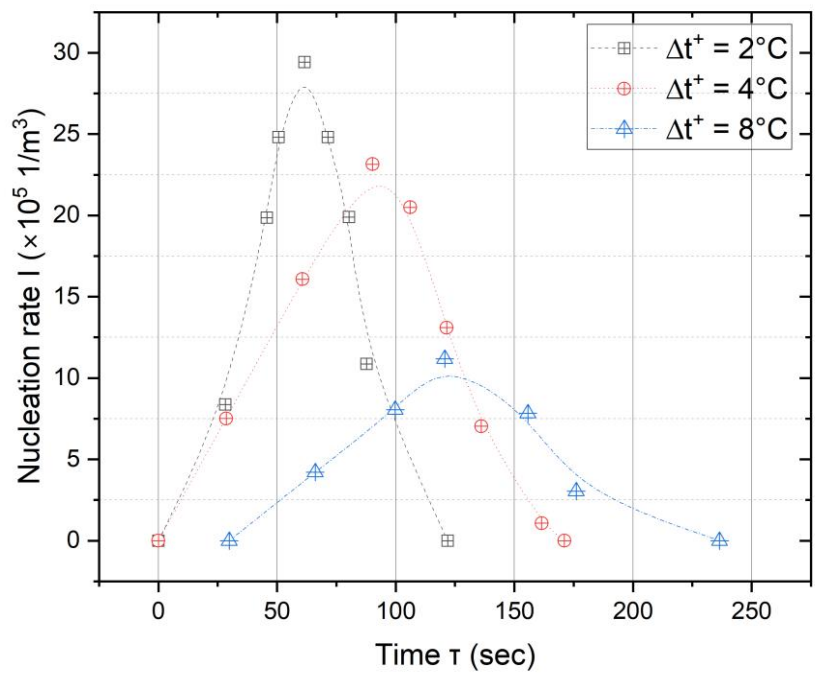

Fig. 3. Change in the rate of formation of camphene nuclei in flat samples depending on the melt overheating temperature.

The rate of non-stationary crystals nucleation ($I_n$) can be described by the following equation [14]:

$$I_n = I \cdot e^{\frac{-\tau_r}{t}}, \qquad (2)$$

where *I* – the speed of stationary nucleation, $\tau_r$ is the relaxation time, and *t* is the time of the nucleus formation process observation.

The logarithm of Eq. (2) gives a linear relationship between $\ln I_n$ and $\tau_r/t$:

$$\ln(I_n) = \ln(I) - \frac{\tau_r}{t}. \tag{3}$$

Appropriate calculations based on this dependence using experimental data (Fig. 3) had shown that, at least for the values of the ascending branches of the function *I(t)*, the obtained results are linear, which indicates a heterogeneous mechanism of the formation of nuclei in flat samples of the studied environments (Fig. 4) [15-16].

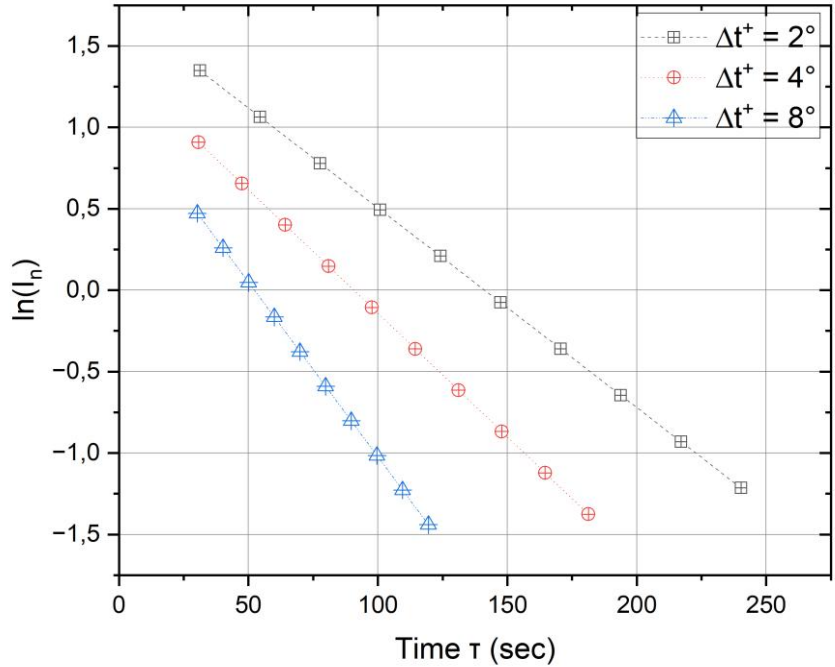

Fig. 4. The change of parameter ln(*I$_n$*) in time *t* for flat samples of camphene at the same level of their supercooling ($\Delta t^- = 10°$) depending on overheating:

A consequence of the change in the rate of formation of nuclei *I* in experimental environments over time (*t*) there is a significant difference in the dispersion in size of crystals (Fig. 5). As we can see, the parameter *I* value growth leads to the dispersion of macrograins sizes growth in 10 times. The same dependencies of the macrostructure on the rate of nucleation were obtained for diphenylamine.

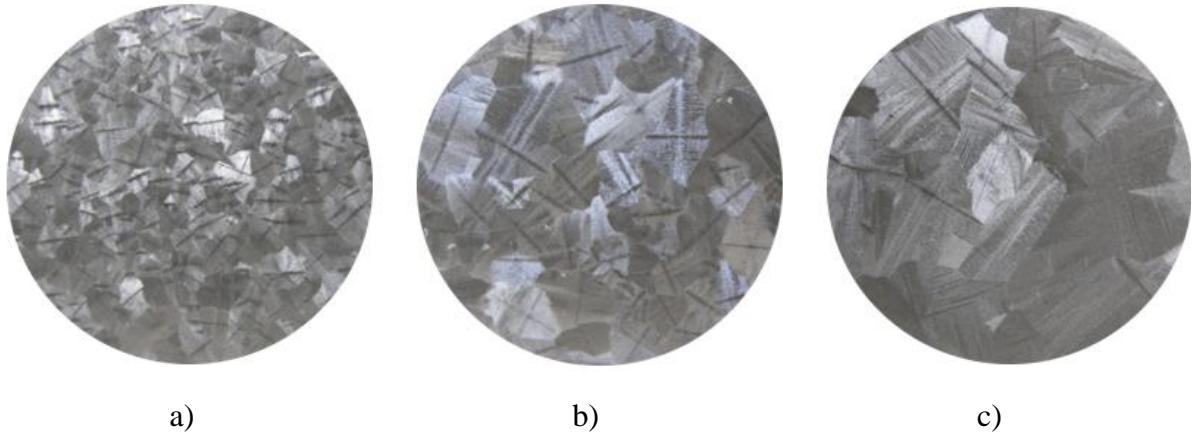

a)   b)   c)

Fig. 5. Macrostructure of camphene depending on the rate of nucleation (optical microscopy): (a) $I = 29 \times 10^5$ 1/m³; (b) $I = 14 \times 10^5$ 1/m³; (c) $I = 6 \times 10^5$ 1/m³

Thus, these experiments on flat samples of small thickness from transparent organic media clearly showed that the dispersion in size of crystals directly depends on the values of nucleation rates at the interphase boundary between the melt and the wall of the mold (model). In turn, the activity of the surfaces of the model walls is essentially determined by the degree of overheating of the melts of the experimental media above the liquidus temperature and the level of their subcooling.

Somewhat different results were obtained by us during the crystallization of the investigated model media in experiments with bulk samples, in which the level of their supercooling values before crystallization was used as a comparative criterion for analysing the nature of nucleation ($\Delta t^-$). Therefore, while the bulk samples of Wood's metal alloy showed good reproducibility of supercooling values before crystallization (± 1°) within the considered superheat temperatures (Fig. 6, line 1), in the bulk samples of transparent organic media of camphene and diphenylamine a wide spread (up to 5 times) of supercooling values before crystallization was recorded ($\Delta t^-$) depending on the number of experiments (Fig. 6, curves 2 and 3). It was previously established that such widespread supercooling for these environments depended on their moisture saturation (they are hygroscopic substances) and other limitedly soluble impurities [14]. At the same time, it was also established that for camphene with each subsequent experiment the value $\Delta t^-$ grew at a slowing rate (Fig. 6, curve 2), and for diphenylamine, it initially increased to a certain maximum and then began to slowly decrease (Fig. 6, curve 3).

Increasing supercooling of camphene $\Delta t^-$ (according to $N$) is explained in assuming that with each subsequent experiment, the moisture content in the test sample decreases due to evaporation and, accordingly, the homogeneity of the melt increases. Increasing the homogeneity of the melt is the main factor that ensures the reproducibility of the experimental results.

With growth in the number of cycles of diphenylamine supercooling the irreversible changes (chemical transformations, dissociation or dissolution of impurities, etc.) gradually occur in the medium itself take place. Indeed, during experiments with diphenylamine, such a fact was recorded that at temperatures of overheating above a certain level (> 100°) its melt changed colour that is it became noticeably darker. In addition, it was established that the melting point of diphenylamine after its overheating, for example, to 120°, decreased by 1.4° and was 51.6°C (Table 1).

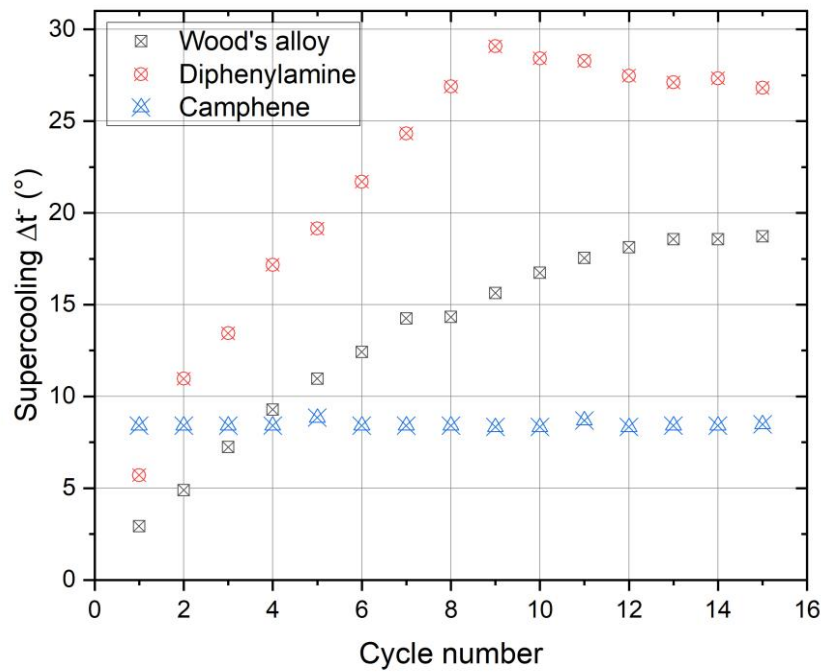

Fig. 6. Dependence of supercooling values of model media before crystallization on the number of repeated tests (sample volume is $18 \times 10^{-6}$ m$^3$). The overheating $\Delta t^+ = 30°$.

Therefore, taking into account the hydrophilicity of organic media, test samples from them before the start of the experiments were subjected to heat treatment at a temperature of 90°C for 60 min., after which the spread of their supercooling values ($\Delta t$) significantly decreased, but remained noticeable. For diphenylamine, the spread of melt supercooling values before crystallization ($\Delta t^-$) for four volumes (3, 6, 12 and $18 \times 10^{-6}$ m$^3$) when overheating $\Delta t^+ = 30°$ averaged 19, 16, 11 and 5%, and for camphene – 17, 14, 9 and 4%, respectively. That is, with an increase in the volume the of samples the scatter of $\Delta t^-$ decreases, which indicates that the mechanism of nucleation in both environments is close to normal, when supercooling before crystallization is determined by many random factors (texture of the mold walls, activity of impurities, temperature and density fluctuations, etc.).

Fig. 7 shows the dependence of the supercooling value ($\Delta t^-$) on the volume ($V$) of experimental samples from model environments. As can be seen for both environments the value of the maximum supercooling with an increase in the volume of the sample ($V$) drops sharply. Then with an excess of a certain value $V$ practically reaches a plateau, i.e. $\Delta t^- \sim const$. The nature of the change in supercooling from the volume in these studies does not confirm the linear dependence of this parameter, predicted by the data of crystallization of microvolumes (droplets) of similar organic media by some authors [15, 16]. This, in our opinion, means that the drops from the experimental media did not have contact with the walls of the form, that is, they were studied in the form of emulsions consisting of dispersed (microscopic) drops distributed in another inert liquid (vaseline oil, glycerin, etc.).

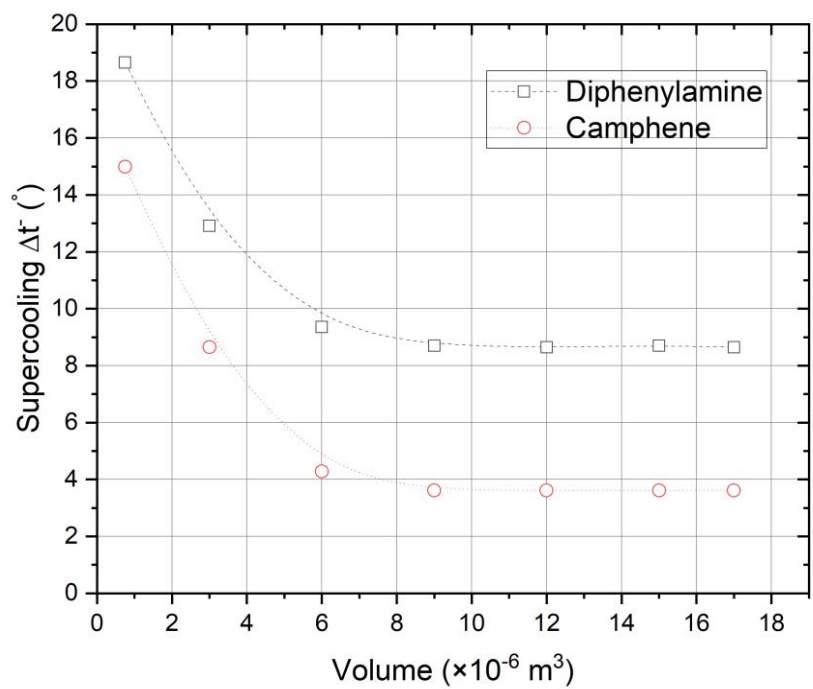

Fig. 7. Dependence of the supercooling values on the sample volume.

Crystallization of macroscopic volumes of any liquid occurs, as a rule, upon contact with the walls of molds and other solid particles, which are always present in them. Accordingly, studying the processes of crystallization of macroscopic volumes is a difficult problem and is of greater scientific and applied interest. Therefore, the dependence of the supercooling of the melts of the model media on the ratio of the area of the wetted surface of the sample ($S$) to its volume ($V$) was determined. The parameter ratio $S/V$ changed by placing test samples in test tubes with different diameters. It was established that with a decrease in the ratio *S/V* for both environments, the maximum supercooling first decreases smoothly, then sharply drops to a certain constant value (Fig. 8). Obviously, this is because with an increase in the volume of the model medium, the probability of the presence of a more active impurity in it increases, which can become the center of crystallization with less supercooling for the given conditions of the experiment. This character of the presented results (Figs. 7 and 8) indicates a heterogeneous mechanism of nucleation in experimental bulk samples, the main feature of which is their relatively low reproducibility (in this case, we are talking about the reproducibility of supercooling values before crystallization).

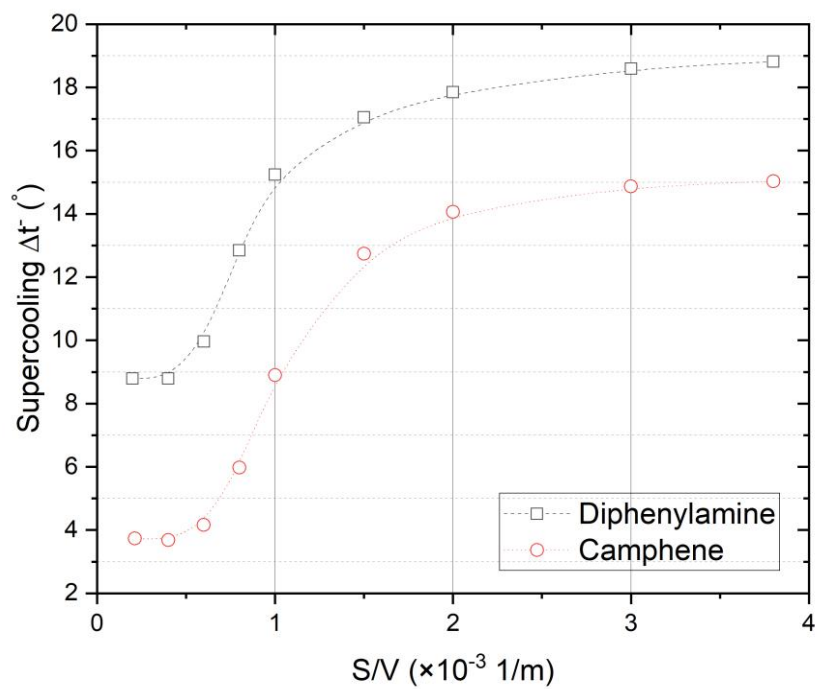

Fig. 8. Dependence of extreme supercooling on the ratio of the surface area to volume wetted by melts.

**Conclusions**

As a result of the work carried out, original physical modelling techniques were developed for the study of crystallization processes and the formation of the structure of metal blanks depending on their scale factors. The experiments on flat samples from transparent organic media have proven that the dispersion in crystals sizes of cast metal blanks with small thickness will depend mainly on the values of nucleation rates at the melt-solid substrate interphase boundary. The role of solid substrates in real alloys is performed by various solid impurities and the walls of casting molds, the activity of which in terms of nucleation, in turn, will be significantly determined by the degree of overheating of metal alloy melts above the liquidus temperature and the level of their supercooling below the solidus temperature.

Also, research demonstrated that in flat samples, the rate of nuclei formation directly depends on the activity of the mold wall surface. This surface activity is significantly determined by the degree of melt overheating above the liquidus temperature and subcooling level. For bulk samples, the dependence of supercooling on overheating is most likely caused by factors such as impurity activity, temperature, and density fluctuations.

The results of the study of crystallization processes in bulk samples of model media indicate that, in contrast to thin samples of small thickness, the rate of nucleation in them is affected by many other large-scale factors, in addition to the activity of the casting molds wall surfaces. Accordingly, it can be asserted that the dispersion of the structures of cast metal blanks of large cross-section, along with the activity of the walls of the casting molds, the temperatures of overheating and cooling of molten metals, will also be influenced by the state of limitedly soluble impurities, convection, temperature and density fluctuations, etc. From these results, it follows that by changing such parameters of melting and pouring of metal alloys as overheating temperature, time of exposure in the superheated state, heating and cooling rates, it is possible to control the degree of supercooling of metal melts before their crystallization (i.e., the rate of crystal formation), respectively, to influence the dispersion of the structures of cast blanks, which implies an influence on the physical and mechanical properties of the cast metal.

The results of the study of crystallization processes in bulk samples indicate that, unlike thin samples of small thickness, the nucleation rate in them, in addition to the activity of the surfaces of the mold walls, other large-scale factors is affected. Accordingly, it can be argued that the dispersion of the structures of cast metal blanks of large cross-section, along with the activity of the walls of the molds, which depends on the overheating and cooling temperatures of molten metals, will also be affected by the state of limitedly soluble impurities, convection, temperature and density fluctuations, etc. These results suggest that by changing such parameters of melting and casting of metal alloys as the superheating temperature, holding time in the superheated state, and heating and cooling rates, it is possible to control the degree of supercooling of metal melts before the onset of their crystallization (i.e., the rate of crystal formation), and accordingly influence the dispersion of the structures of cast blanks, which implies an impact on the physical and mechanical properties of the cast metal.